\documentclass[aps,prb,10pt,twocolumn,superscriptaddress]{revtex4-2}
\usepackage{amsmath}
\usepackage{amssymb}
\usepackage{graphicx}
\usepackage{dcolumn}
\usepackage{bm}
\usepackage[utf8]{inputenc}
\usepackage{verbatim}
\usepackage{color}
\definecolor{lblu}{rgb}{0, 0.5, 0.65}
\begin{document}
\title{Systematic manipulation of the surface conductivity of SmB$_6$}

\author{M. Victoria Ale Crivillero}
\affiliation{Max-Planck-Institute for Chemical Physics of Solids,
N\"othnitzer Str. 40, 01187 Dresden, Germany}
\author{M. K\"{o}nig}
\affiliation{Max-Planck-Institute for Chemical Physics of Solids,
N\"othnitzer Str. 40, 01187 Dresden, Germany}
\author{J. C. Souza}
\affiliation{Instituto de F\'{i}sica ``Gleb Wataghin'', UNICAMP, 13083-859
Campinas, S\~{a}o Paulo, Brazil}
\affiliation{Max-Planck-Institute for Chemical Physics of Solids,
N\"othnitzer Str. 40, 01187 Dresden, Germany}
\author{P. G. Pagliuso}
\affiliation{Instituto de F\'{i}sica ``Gleb Wataghin'', UNICAMP, 13083-859
Campinas, S\~{a}o Paulo, Brazil}
\author{J.~Sichelschmidt}
\affiliation{Max-Planck-Institute for Chemical Physics of Solids,
N\"othnitzer Str. 40, 01187 Dresden, Germany}
\author{Priscila F. S. Rosa}
\affiliation{Los Alamos National Laboratory, Los Alamos, NM 87545, USA}
\author{Z. Fisk}
\affiliation{Department of Physics, University of California,
Irvine, California 92697, USA}
\author{S. Wirth}
\email{Steffen.Wirth@cpfs.mpg.de}
\affiliation{Max-Planck-Institute for Chemical Physics of Solids,
N\"othnitzer Str. 40, 01187 Dresden, Germany}
\date{\today}

\begin{abstract}
We show that the resistivity plateau of SmB$_6$ at low temperature, typically
taken as a hallmark of its conducting surface state, can systematically be
influenced by different surface treatments. We investigate the effect of 
inflicting an increasing number of hand-made scratches and microscopically
defined focused ion beam-cut trenches on the surfaces of flux-grown
Sm$_{1-x}$Gd$_x$B$_6$ with $x =$ 0, 0.0002. Both treatments increase the 
resistance of the low-temperature plateau, whereas the bulk resistance at
higher temperature largely remains unaffected. Notably, the temperature at
which the resistance deviates from the thermally activated behavior decreases
with cumulative surface damage. These features are more pronounced for the
focused ion beam treated samples, with the difference likely being related to
the absence of microscopic defects like subsurface cracks. Therefore, our
method presents a systematic way of controlling the surface conductance.
\end{abstract}
\maketitle   

\section{Introduction}
Over the past decade, the proposed topological Kondo insulator SmB$_6$ has
seen a surge of research interest \cite{li20} despite its more than
half-a-century-old history \cite{vai64,men69}. This interest stems from a
combination of complex correlated electron physics and the proposed
topologically non-trivial surface states resulting from spin-orbit-driven
band inversion in the bulk \cite{dze10,tak11,dze16}.

Irrespective of the direct involvement of the surface in the topical physics,
relatively little is known about its properties. While the bulk of SmB$_6$ is
known for its intermediate and temperature-dependent Sm valence of
approximately 2.6 at low temperature \cite{vai64,tar80,miz09,lut16,uts17}, the
valence at the surface appears to be closer to 3+ \cite{lut16,uts17}. This
change in valence could be related to the formation of Sm$_2$O$_3$ near the
surface, resulting from an oxidation of the near-surface Sm. A changed surface
chemistry may also shift the chemical potential at the surface \cite{par16}
and may lead to time-dependent surface properties \cite{zhu13}. Consequently,
in numerous studies relying on highly surface-sensitive techniques like
scanning tunneling microscopy and spectroscopy (STM/S) \cite{yee13,roe14,
ruan14,jiao16,jiao18,sun18,pir20} and angle-resolved photoemission
spectroscopy (ARPES) \cite{jia13,fra13,den14,nxu14,sug14,hla18,mat20}, SmB$_6$
surfaces were prepared by {\em in situ} cleaving in ultra-high vacuum (UHV)
conditions. However, SmB$_6$ is difficult to break, and surfaces perpendicular
to the main crystallographic axes of the cubic structure (space group
$Pm\bar{3}m$) are polar in nature, giving often rise to (2$\times$1)
reconstructed surfaces. Notably, even cleaved surfaces may exhibit valence
inhomogeneities \cite{zab18} and band-bending effects
\cite{neu13,ish15,mat20}.

Also for cases of less surface-sensitive techniques, like resistivity
measurements, surfaces often need to be prepared, e.g. by polishing or etching
(see e.g. \cite{kim13,wol13,wol15,sye15,bis17,eo18,eo19,fuh19,eo20b}).
However, such surface preparation may influence the surface itself, e.g. by
disrupting the crystal structure at the surface, introducing impurities or,
again, changing the Sm valence. One particularly interesting example here is
the creation of so-called subsurface cracks by rough polishing \cite{eo20}.
These subsurface cracks constitute additional surfaces with their own surface
states, which conduct in parallel to the actual sample surface. Hence, care
has to be taken when comparing different results since differences in the
applied surface preparation procedure may result in differences in the
measured properties. To make things more complicated, there can also be
differences between samples grown by either floating zone or Al flux
technique, not only intrinsically \cite{phe16,tho19,ghe19,ros20,eo20b} but
also with respect to the impact of surface preparation as shown exemplary for
etched surfaces \cite{eo20b}.

We here apply a systematic way of manipulating the sample surface by utilizing
a focused ion beam (FIB), complemented by a rather crude surface scratching.
The low-temperature resistivity plateau of our flux-grown SmB$_6$ samples, typically taken as a hallmark of the conducting surface state, can be
influenced considerably, yet consistently, by both surface treatments,
indicating impaired surface states. As expected, the thermally activated
transport across the bulk gap is not affected significantly by the surface
treatments.

\section{Experimental}
The samples Sm$_{1-x}$Gd$_x$B$_6$ used in this study were grown by the Al
flux technique \cite{rosa18} with Gd content $x =$ 0 and 0.0002. The tiny
amount of Gd for the latter samples was confirmed by magnetic susceptibility
measurements. It allowed electron spin resonance (ESR) measurements, which
will be reported elsewhere \cite{souX}. We did not observe any noticeable
differences in the here reported properties of samples 
with $x =$ 0 and 0.0002. Therefore, we concentrate in the following on samples
$x =$ 0.0002 which were studied more extensively. Energy-dispersive x-ray
(EDX) spectroscopy conducted within our FIB equipment at pressures in the
10$^{-6}$ mbar range did not show any elements other than Sm, B, O and Al,
with Gd being below the detection limit. Upon using the FIB to remove a
layer of a few $\mu$m thickness, the Al signal is no longer detectable within
these sputtered areas.

As a crude way of disrupting the surface we cut lines by means of a diamond
scribe. Because of the hardness of SmB$_6$, considerable force had to be
\begin{figure}[t]
\includegraphics*[width=0.39\textwidth]{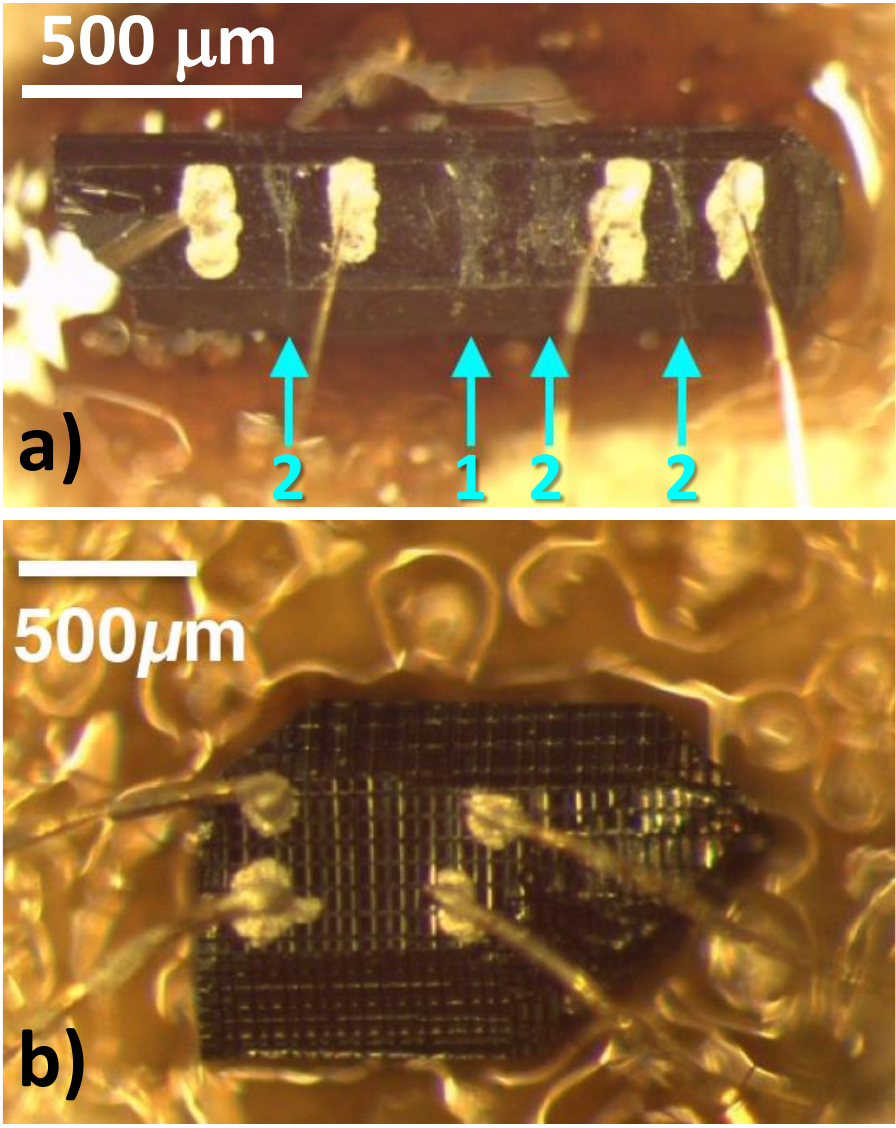}
\caption{Images of two exemplary SmB$_6$ samples. (a) Optical microscopy image
of a surface with scratches between the glued-on contacts. The scratches
inflicted in the first (1) and second (2) run are marked by arrows. (b)
Optical image of a sample with lines cut by FIB (after 6 runs) and with
contacts attached.}   
\label{photo41109}  \end{figure}
applied to inflict the line damage to the sample surface as shown in Fig.\
\ref{photo41109}(a). In this example, the first scratch (marked by an arrow
and a number) was only applied to the front surface. In a second step, more
scratches were applied, and all scratches now cover also back and side
surfaces to form closed rings approximately perpendicular to the long sample
axis. In a third step, the scratches were deepened by applying more force to
the diamond scribe. Figure \ref{photo41109}(a) was taken subsequent to the
second scratching and with contacts for resistance measurements attached.

\begin{figure}[t]
\includegraphics*[width=0.48\textwidth]{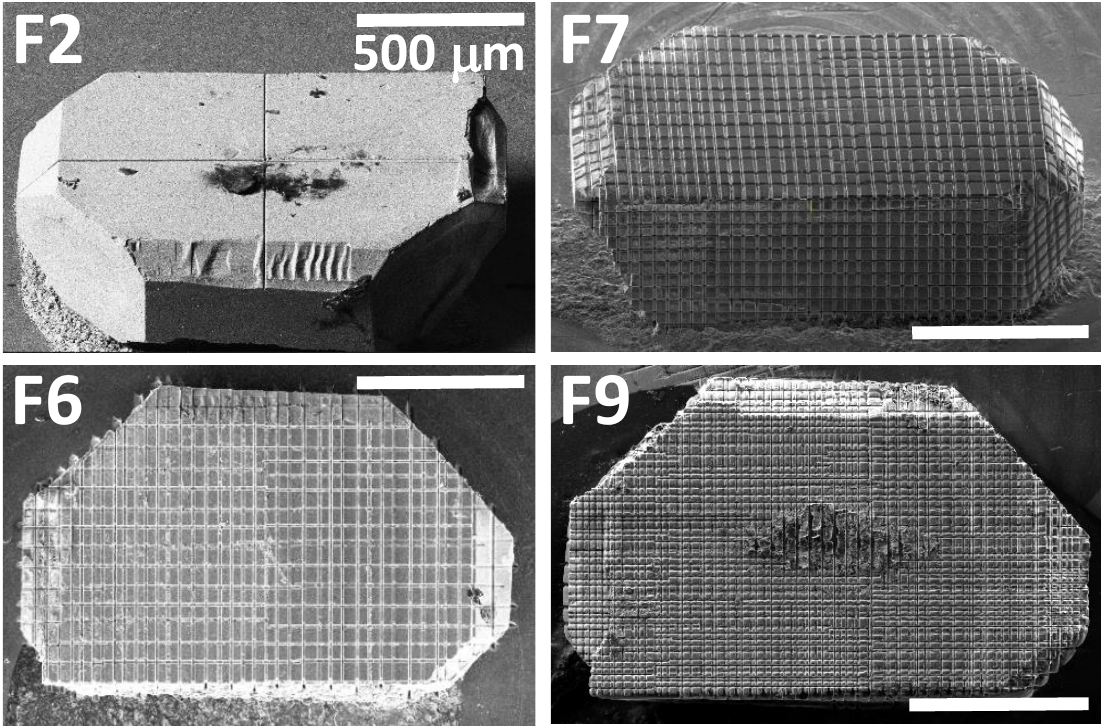}
\caption{SEM images of an SmB$_6$ sample after FIB-cutting an increasing
number of lines. Shown are examples after FIB-runs F2, F6, F7 and F9. The dark
patches at the centers of F2 and F9 are residue of paraffin used for fixing
the sample. All scale bars: 500 $\mu$m.}  
\label{fibbing}  \end{figure}

In an effort to structure the surfaces of our samples in a much more
systematic and controlled fashion, we utilized a FIB. Trenches of about 7--10
$\mu$m in depth were cut by Xe ions at beam currents of 500 nA with
acceleration voltage of 30 kV in consecutive runs. In a first run (denoted
F1 in the following), a single line is cut across the middle of the sample,
dividing the surface in two parts. In a second run (F2), each half is
subdivided into two fields of similar size by cutting a line perpendicular to
the first one; the resulting crossed lines are seen in the scanning
electron microscopy (SEM) image Fig.\ \ref{fibbing}-F2. In subsequent runs,
the number of lines is about doubled in each direction by cutting additional
lines approximately parallel to the already existing ones. The resulting grids
of FIB-cut lines after selected runs are presented in Fig.\ \ref{fibbing}. In
case of this particular sample, during runs F1 -- F6 cuts were inflicted only
on the front and back surface (the front surface is seen in Figs.\
\ref{fibbing}-F6 and -F9). Run F7 was used to transfer the existing grid to
all side surfaces, see Fig.\ \ref{fibbing}-F7. Consecutive runs F8 and F9
included the front and back surfaces. After the final run (F9), this sample
was FIB-cut to a minimum grid line distance of about 15 $\mu$m. We note that
this distance is still large compared to the effective carrier mean free path
$\ell <$ 1 $\mu$m \cite{souX,har18}.

Resistance measurements were usually conducted after some FIB runs or line
scratches, using a Physical Property Measurement System (PPMS) by Quantum
Design, Inc. In case of FIB-cut sample surfaces, van der Pauw-type
measurements were conducted. A sample (different to the one presented in
Fig.\ \ref{fibbing}) after FIB run F6 with contacts attached is shown in
Fig.\ \ref{photo41109}(b).

\section{Results}
\subsection{Sm$_{1-x}$Gd$_x$B$_6$ samples with scratched surfaces}
\begin{figure}[t]
\includegraphics*[width=0.48\textwidth]{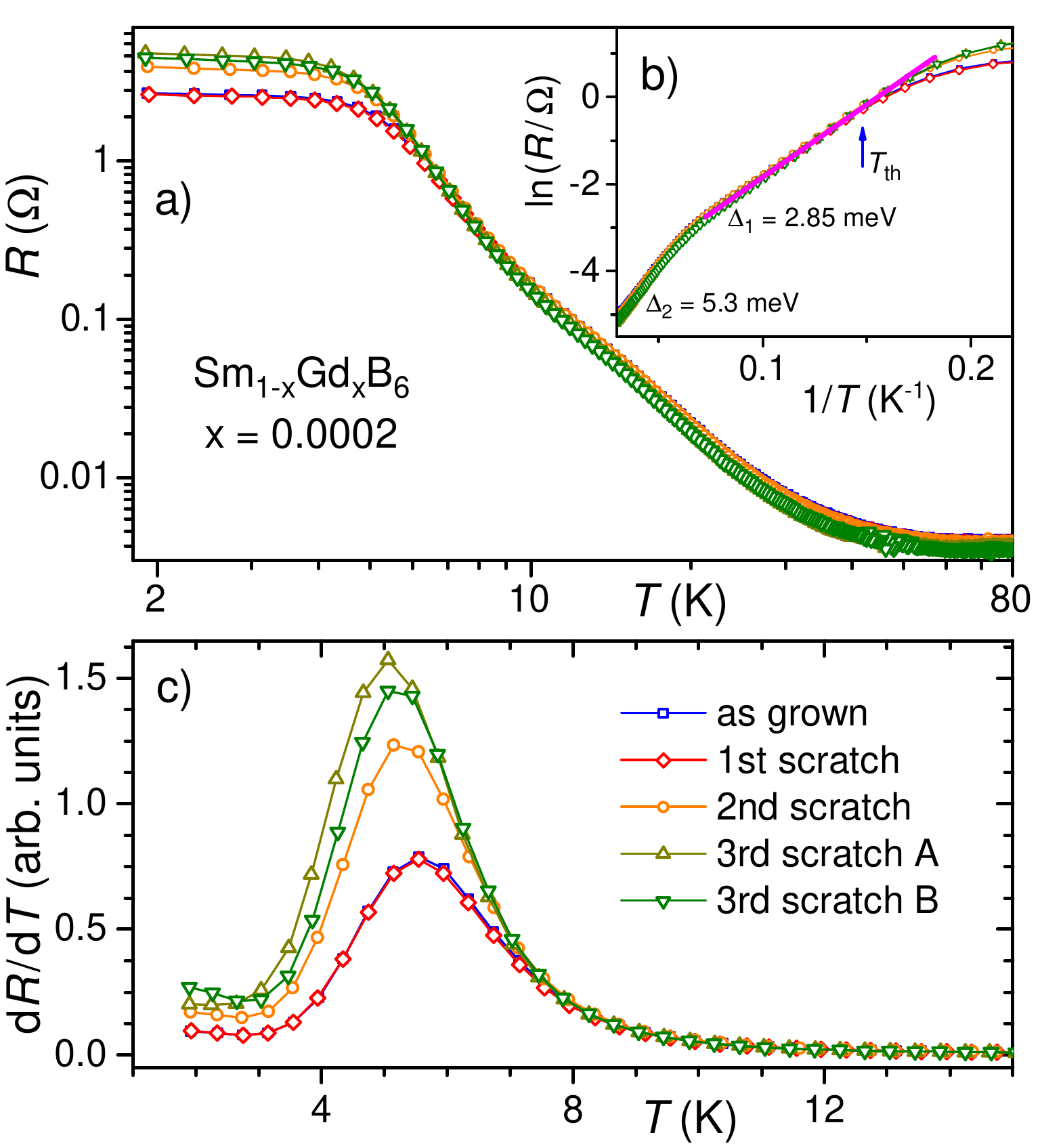}
\caption{(a) Double-logarithmic plot of the resistance of a
Sm$_{1-x}$Gd$_x$B$_6$ sample with $x =$ 0.0002 before, and subsequent to, an
increasing number of scratches. After the third scratching, resistances were
measured through two sets of contacts, A and B. (b) $\ln R$ vs. $1/T$
representation of the intermediate temperature data. The magenta line
illustrates thermally activated behavior, from which $R(T)$ deviates at
$T_{\rm th}$ (shown for the as-grown sample). (c) Derivative of the low-$T$
resistance as a function of temperature.}  \label{rscrat}
\end{figure}

Resistances of the sample shown in Fig.\ \ref{photo41109}(a) before and
after inflicting an increasing number of scratches to the sample surface are
presented in Fig.\ \ref{rscrat}(a). Clearly, the first scratch did not
significantly change the resistance, possibly because the first scratch did
not form a closed ring around the sample. Consecutive scratches formed closed
rings and introduced resistance changes. These changes, however, are
exclusively limited to the low-temperature regime, as shown in the different
representations in Fig.\ \ref{rscrat}(a)-(c). This finding is in agreement
with the fact that the resistance at higher $T$ reflects bulk properties while
the surface state dictates the resistance behavior only at $T$ below a few K.
The bulk hybridization gap $\Delta$ can be estimated from $R(T) \propto \exp
(\Delta /k_{\rm B} T)$, where $k_{\rm B}$ is the Boltzmann constant.
Typically, for pure \cite{fla01,eo19} and slightly Gd-substituted 
\cite{jiao18,sou20} SmB$_6$ two regimes with different gap values are observed
depending on the $T$-range considered. This also holds for our measurements
with $\Delta_1 =$ 2.85($\pm 0.07$) meV and $\Delta_2 =$ 5.3($\pm 0.1$) meV,
independent of the scratches, see Fig.\ \ref{rscrat}(b). However, the
scratches do influence the lower bound $T_{\rm th}$ of the temperature range
within which $R(T)$ can be described by thermally activated behavior (the
latter is marked by a magenta line in Fig.\ \ref{rscrat}(b)). Obviously, there
is a clear trend: the more pronounced the scratches, the lower $T_{\rm th}$.
We find for the as-grown sample and after the first scratch $T_{\rm th}
\approx$ 7~K [see arrow in Fig.\ \ref{rscrat}(b)], after the second scratching
$T_{\rm th} \approx$ 6.4 K, and $T_{\rm th} \approx$ 6.1 K after the third
scratching. This trend is also seen in the derivative d$R$/d$T$ in Fig.\
\ref{rscrat}(c). As outlined in Ref.\ \onlinecite{eo19}, the thermally
activated behavior, i.e. the exponential increase of $R(T)$, is a clear
hallmark of the bulk resistance, which is superseded by the {\em additional}
surface component upon lowering $T$ assuming a parallel conductance model
\cite{wol13,sye15,lee16}. The low-$T$ resistance plateau indicates the
presence of the surface states even after scratching. Yet, based on the trend
of $T_{\rm th}$, the crossover from bulk-dominated conductivity (roughly above
10~K) to surface-dominated conductivity (below about 3 K) appears to take
place at lower temperature. The increased value of the low-$T$ resistance
plateau measured on the damaged surfaces could be caused by either a decreased
conductivity of the intrinsic surface state or an additional damage layer at
the scratched areas below which the intrinsic surface state reconstructs (or
a combination thereof). However, the surface state still develops and appears
to govern the $R(T)$ below a similar $T \approx$ 3 K at which the $R(T)$-slope
does no longer change, see Fig.\ \ref{rscrat}(c). Here we note that subsequent
to measurement ``3rd scratch A'' the contacts were completely removed and
attached anew for measurement ``B'', showing that the contacts themselves have
no significant influence on $R(T)$. In particular, the difference of the
$R(T)$-values at low $T$ is less than 7\% (compared to $\sim 20$\% change
between 2nd and 3th scratch).

\subsection{FIB-cut trenches on sample surfaces}
\begin{figure}[t]
\includegraphics*[width=0.48\textwidth]{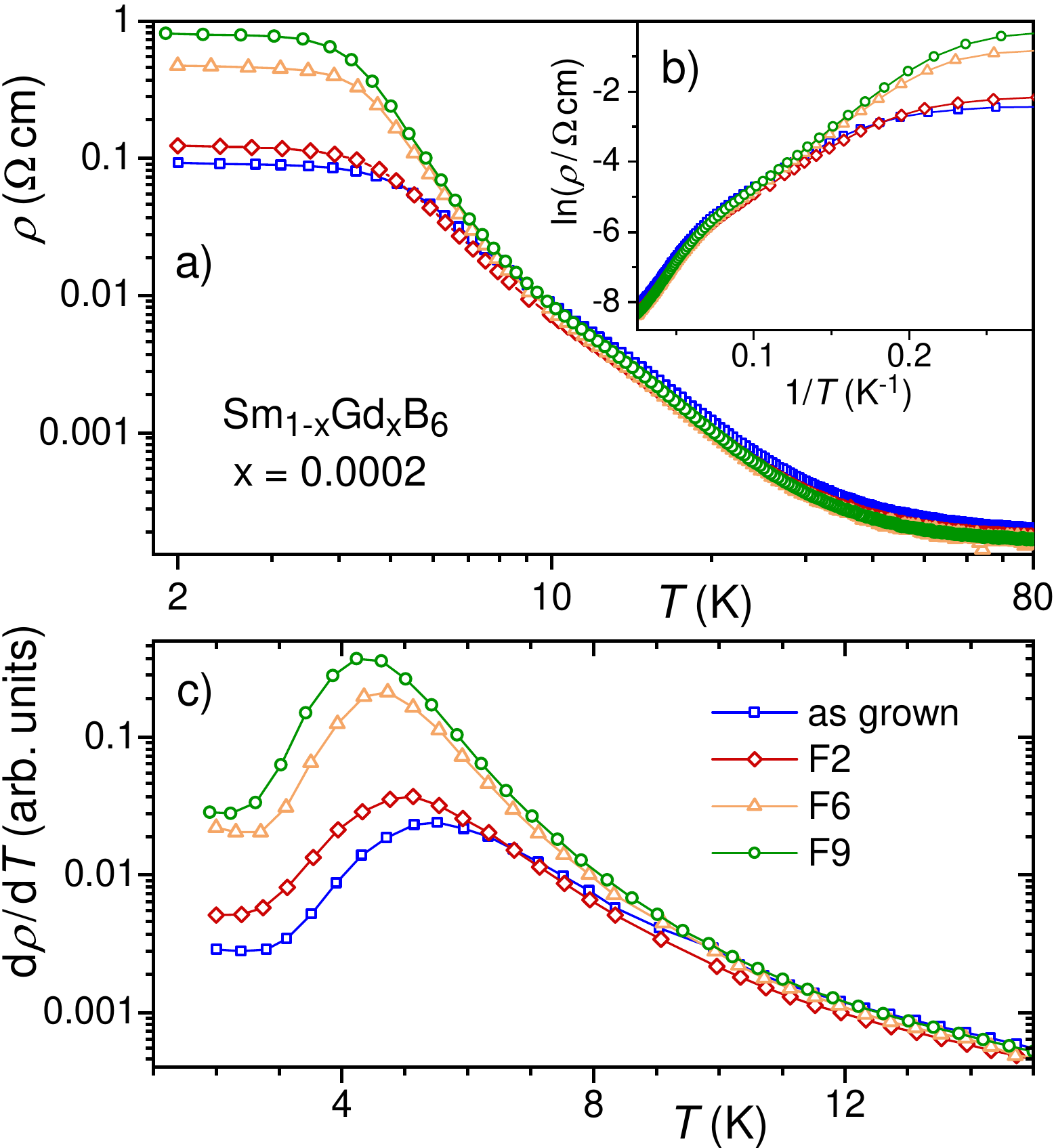}
\caption{(a) Double-logarithmic plot of the resistivity $\rho$ of
Sm$_{1-x}$Gd$_x$B$_6$ with $x =$ 0.0002 before and after FIB-cutting an
increasing number of lines, see Fig.\ \ref{fibbing}. (b) $\ln \rho$ vs. $1/T$
representation of the intermediate temperature data. (c) Low-$T$ range of the
derivative d$\rho /{\rm d}T$.} \label{rfib}
\end{figure}

In order to manipulate the sample surface in a much more controlled and
systematic way, we also measured the resistivity of FIB-treated samples as
described above. Exemplary resistivity data of a sample Sm$_{1-x}$Gd$_x$B$_6$
with $x =$ 0.0002 are given in Fig.\ \ref{rfib}(a) for an increasing number
of FIB-cut lines on its surfaces. We note that the contacts needed to be
removed before each subsequent FIB run. Albeit great care was taken to
re-attach the contacts after the FIB run at the very same positions, a
marginal influence on the resistivity values cannot be excluded entirely.
Due to the small size and position of the contacts, there was no conducting
path between contacts uninterrupted by FIB-cut lines -- even via side surfaces
-- after FIB-run F5 already [as can be inferred from Fig.\ \ref{photo41109}(b)
for the second sample].

Already the first cross of FIB-cut lines (F2) increases the $\rho (T)$-values
within the plateau at low $T$ compared to the as-grown surface by more than
30\%, which appears to be well beyond the geometry inaccuracy. Interestingly,
the reduction of $T_{\rm th}$ after FIB-run F2 is very similar to those after
the second and third scratch, i.e. for comparable number of lines/scratches.
Upon increasing the FIB-cut line density, the low-$T$ resistivity increases
further such that $\rho (T)$ of F9 at low temperature exceeds the value of the
as-grown surface by almost an order of magnitude. Just as pronounced is the
concomitant drop of $T_{\rm th}$ by 2 K from as grown to F9, see Tab.\
\ref{tabFIB}. Most other parameters remain largely unaffected by the FIB
surface structuring, only $\Delta_1$ appears to be slightly modified.
Nonetheless, the bulk sample properties remain essentially unaltered by the
FIB treatment. Very similar trends were observed on a second FIB-cut sample,
shown in Fig.\ \ref{photo41109}(b).

The increased resistivities upon damaging the sample surfaces is in contrast
to its observed decrease for substituted or intentionally imperfect samples \cite{fla01b,phe14,kan16,ore17,jiao18,dem20} or ion irradiated samples
\cite{wak15}. This might imply that our surface treatments by FIB or
scratching do not influence the surfaces on the whole, but rather act on the
surface states locally. On the other hand, an increased slope of the low-$T$
resistivity appears to generally indicate a diminished surface state.

\section{Discussion}
\begin{figure}[t]
\includegraphics*[width=0.4\textwidth]{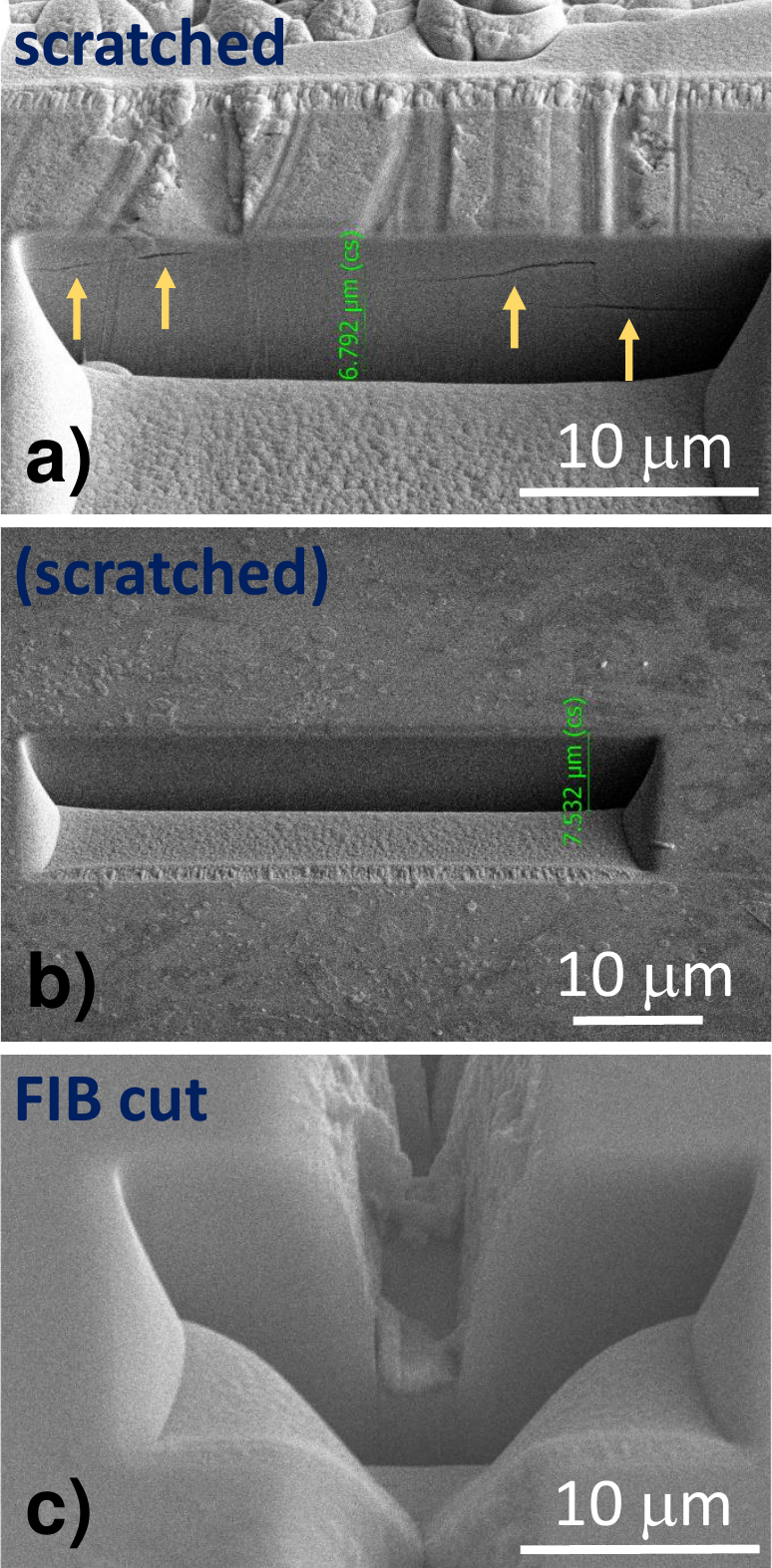}
\caption{Exemplary SEM images of (a) a sample cross section underneath a
scratched line. Cracks are marked by arrows. (b) Same sample as in (a) but
cross section taken at an area away from any scratch. (c) Cross section
surrounding a FIB cut line.} \label{sem}
\end{figure}

In Ref.\ \onlinecite{eo20}, the influence of subsurface cracks on the total
resistivity is discussed. Such subsurface cracks provide additional conduction
channels, and a decreased low-$T$ resistance upon surface scratching was
reported \cite{eo20}. Subsurface cracks could also be found below our
scratched surface areas, see the arrow marks in Fig.\ \ref{sem}(a). We note
that such subsurface cracks could so far be found exclusively underneath the
scratches, presumably indicating that the unscratched sample regions
\begin{table}[t]
\caption{Fit parameters for the two thermally activated gaps of the FIB-cut
sample shown in Fig.\ \ref{fibbing}, obtained from the resistivity data of
Fig.\ \ref{rfib}. The gap values $\Delta$ and the temperature ranges within
which the fits hold are given. $d$ denotes the approximate distance between
FIB-cut lines (or the lines and the sample perimeter in case of F2).}
\label{tabFIB}
\begin{ruledtabular}
\begin{tabular}{c|c|ccc|cc}
sample & & \multicolumn{3}{c|}{gap at lower $T$} & \multicolumn{2}{c}{gap
at higher $T$}\\ \hline
\rule{0pt}{10pt}& $d$ & $\Delta_1$ & $T_{\rm th}$ & upper $T$ & $\Delta_2$ &
fitted $T$-range\\
 & $\mu$m & meV     & K       & K         & meV       & K \\ \hline
\rule{-3pt}{10pt}
as grown &     & 2.4  & 7.1  & 12.3       & 5.2       & 18.4 -- 29.2 \\
F2       & 700 & 2.4  & 6.3  & 12.3       & 5.2       & 18.7 -- 29.4 \\
F6       & 40  & 2.8  & 5.4  & 12.3       & 5.3       & 18.8 -- 30.3 \\
F9       & 15  & 2.9  & 5.1  & 12.4       & 5.4       & 18.6 -- 29.4 \\
\hline \rule{-3pt}{10pt}
error &  & $\pm 0.2$ & $\pm 0.3$ & $\pm 0.4$  & $\pm 0.2$ & $\pm 0.8$
\end{tabular}
\end{ruledtabular}
\end{table}
(including the pristine samples) are free of such subsurface cracks, Fig.\
\ref{sem}(b). The subsurface cracks can be found down to a few micrometers
below the scratched surface. However, in contrast to the earlier findings
\cite{eo20} the sample resistance {\em increases} with scratching in our case,
Fig.\ \ref{rscrat}(a). Here we recall that our scratches encircle the whole
surface without leaving any possible current path on the surface untouched.
Therefore, our approach seems to emphasize the impact of the surface
conductance to the total sample resistance compared to Corbino-type
measurements \cite{wol15,eo19,eo20}. We therefore infer that the value of the
low-$T$ resistance plateau is the result of two counteracting effects: while
the subsurface cracks lower this value by introducing additional conductance
channels, the surface conductance itself is hampered due to scratching, as
also indicated by the lower $T_{\rm th}$ values. In this respect, the
intermediate and high temperature resistance regime provides important hints
of largely unchanged bulk properties.

This picture is corroborated by the results of the FIB-cut samples. Albeit
the trenches inflicted by the FIB cut deeper into the sample (up to about 10
$\mu$m) compared to the scratches (typically a few micrometers, with depths
up to about 5 $\mu$m), we did so far not find any indication for subsurface
cracks on FIB-treated samples, see example in Fig.\ \ref{sem}(c). The
material directly at the bottom of the FIB-cut trenches, and to a lesser
extent also at the sidewalls, is typically turned amorphous to a depth of
several tens of nanometers and, in case of preferential sputtering,
non-stoichiometric \cite{mol18}. Below this affected layer, the crystal
structure is usually well preserved, with only occasional lattice defects
caused by the ion bombardment. In this sense, the FIB treatment can be
considered a controlled and systematic way of manipulating the surface
conductance. The low-$T$ saturation of $\rho (T)$, Fig.\ \ref{rfib}(a),
indicates that the conducting surface layer, albeit possibly encumbered, is
still subsisting. We note that in one case we also conducted an abrasion of
the whole sample surface of about 3 $\mu$m deep by rastering the entire
sample surface with the ion beam in a last run (i.e.\ after FIB-cutting a
line grid) and still observed indications of the surface layer, in line with
Ref.\ \onlinecite{miy17}. This finding has an interesting consequence:
The above-mentioned amorphous layer then covers the whole sample surface.
Since this layer very likely prevents a surface reconstruction from forming,
we can in all likelihood rule out a $2 \times 1$ surface reconstruction (as,
e.g., observed in some cases by STM \cite{yee13,roe14,pir20}) causing the
conducting surface states. Also, the polarity change at the interface between
amorphous and crystalline SmB$_6$ is certainly smaller compared to a pure
SmB$_6$ surface. In consequence, all this makes conducting surface states
driven by a non-trivial topology of the crystalline SmB$_6$ more likely.

There are at least two contributions which may cause the increase of $\rho(T)$
at low-$T$ upon FIB-cutting trenches: i) The surface state may be tampered
with and ii) the surface area increases with the number of lines. The latter,
however, appears not to be a decisive factor as an increased $\rho (T)$ is
already observed for a small number of lines. As an example, the surface area
of surface F6 in Fig.\ \ref{rfib} is less than doubled compared to the as-grown
one, but $\rho$(2~K) increased by a factor of more than 5. This strong increase
of $\rho (T)$ at low temperature, along with the concomitantly lowered
$T_{\rm th}$ (Tab.\ \ref{tabFIB}) as well as preliminary ESR results
\cite{souX} suggest a reduction of the contribution of the surface states to
the sample conductivity, possibly due to an FIB-induced depletion of the
surface states. This might be related to a confinement of the surface states.
In addition, disorder effects (in the bulk or/and near the surface due to our
treatments) can be important in Kondo insulators as disorder can greatly
affect the hybridization gap \cite{sen20,abe20} and, in turn, the surface
states. As one example, the Sm valence near the surface can be modified
\cite{zab18,fuh19}, which could introduce changes to the surface conductivity.

As mentioned above, it is unlikely, and also not seen in our attempts, that
the FIB treatment induces subsurface cracks. On the other hand, similar to
the case of scratched sample surfaces, the surface state appears to be
tampered with. Therefore, one may speculate that the relatively small increase
of the low-$T$ resistance for the scratched surfaces compared to the
FIB-treated ones is related to the subsurface cracks in the former. Of course,
the severity of the inflicted damage to the respective surface may also
differ.

\section{Conclusion}
We showed that introducing localized damage to the SmB$_6$ surface by
different treatments, like mechanical surface scratching and FIB-cut trenches,
can alter the low-temperature resistance plateau significantly. We find that
the measured low-temperature $R$-value depends sensitively on the type of
surface treatment and the structural damage incurred. In our cases, the bulk
resistivity at higher temperature remains largely unchanged and hence, the
ratio between the resistances at high and at lowest temperature is not a good
measure of the sample quality. However, the low-temperature limit to which
the resistance follows a thermally activated behavior is found to be related
to the severity of damage inflicted to the surface.

More generally, the systematic and well-controlled surface treatment by FIB
as presented here may provide a path for modification and patterning of
surface states as recently suggested theoretically \cite{sac15}.

\section*{Acknowledgments}
JCS acknowledges support through the RIA Doctorate funding by FAPESP, Brazil
(Grants no. 2020/12283-0, 2018/11364-7, 2017/10581-1). Work at Los Alamos
National Laboratory was performed under the auspices of the U.S. Department
of Energy, Office of Basic Energy Sciences, Division of Materials Science and
Engineering. ZF acknowledges support from the LANL G.\ T.\ Seaborg Institute.

\end{document}